\documentclass[aps,showpacs,tightenlines,twocolumn,nofootinbib,nobibnotes,superscriptaddress]{revtex4-1}
\usepackage{amsmath,amssymb,amsfonts,bm}
\usepackage{graphicx}
\usepackage{epstopdf}
\usepackage{dcolumn}
\usepackage{mathrsfs}
\usepackage{tgtermes}
\usepackage[colorlinks=true,linkcolor=red,citecolor=blue, urlcolor=blue,bookmarks=false]{hyperref}
\bibpunct{[}{]}{,}{n}{}{}
\usepackage{verbatim}
\usepackage{amsthm,amsmath,amssymb}
\usepackage{mathrsfs}
\usepackage{booktabs}
\usepackage{multirow}

\begin{document}
\title{Higher-order topological insulators in hyperbolic lattices}
\date{\today }
\author{Zheng-Rong Liu}
\affiliation{Department of Physics, Hubei University, Wuhan 430062, China}
\author{Chun-Bo Hua}\email{chunbohua@hbust.edu.cn}
\affiliation{School of Electronic and Information Engineering, Hubei University of Science and Technology, Xianning 437100, China}
\author{Tan Peng}
\affiliation{School of Mathematics, Physics and Optoelectronic Engineering, Hubei University of Automotive Technology, Shiyan 442002, China}
\author{Rui Chen}\email{chenr@hubu.edu.cn}
\affiliation{Department of Physics, Hubei University, Wuhan 430062, China}
\affiliation{Department of Physics, The University of Hong Kong, Pokfulam Road, Hong Kong 999077, China}
\author{Bin Zhou}\email{binzhou@hubu.edu.cn}
\affiliation{Department of Physics, Hubei University, Wuhan 430062, China}

\begin{abstract}
 To explore the non-Euclidean generalization of higher-order topological phenomena, we construct a higher-order topological insulator model in hyperbolic lattices by breaking the time-reversal symmetry (TRS) of quantum spin Hall insulators. We investigate three kinds of hyperbolic lattices, i.e., hyperbolic $\{4,5\}$, $\{8,3\}$ and $\{12,3\}$ lattices, respectively. The non-Euclidean higher-order topological behavior is characterized by zero-energy effective corner states appearing in hyperbolic lattices. By adjusting the variation period of the TRS breaking term, we obtain 4, 8 and 12 zero-energy effective corner states in these three different hyperbolic lattices, respectively. It is found that the number of zero-energy effective corner states of a hyperbolic lattice depends on the variation period of the TRS breaking term. The real-space quadrupole moment is employed to characterize the higher-order topology of the hyperbolic lattice with four zero-energy effective corner states. Via symmetry analysis, it is confirmed that the hyperbolic zero-energy effective corner states are protected by the particle-hole symmetry $P$, the effective chiral symmetry $Sm_{z}$, and combined symmetries $C_{p}T$ and $C_{p}m_{z}$. The hyperbolic zero-energy effective corner states remain stable unless these four symmetries are broken simultaneously. The topological nature of hyperbolic zero-energy effective corner states is further confirmed by checking the robustness of the zero-energy modes in the hyperbolic lattices in the presence of disorder. Our paper provides a route for research on hyperbolic higher-order topological insulators in non-Euclidean geometric systems.
\end{abstract}

\maketitle

\section{Introduction}

The study of topological insulators is an important field in condensed matter physics~\cite{RevModPhys.82.3045, RevModPhys.83.1057, RevModPhys.88.021004, RevModPhys.89.040502, RevModPhys.89.041004, Wlfle2018}. Topological insulators in the $d$-dimensional geometry host a $d$-dimensional gapped bulk and $(d-1)$-dimensional gapless boundary states~\cite{PhysRevLett.61.2015, PhysRevLett.95.146802, PhysRevLett.95.226801, doi:10.1126/science.1133734, doi:10.1126/science.1148047, PhysRevX.7.041069}. These boundary states are protected by the topological properties of the bulk, characterized by a quantized number called the topological invariant~\cite{PhysRevLett.49.405, doi:10.1098/rspa.1984.0023, PhysRevLett.95.226801, RevModPhys.82.3045, PhysRevA.78.033834}. Recently, the notion of topological insulators has been generalized, and a class of topological materials called higher-order topological insulators (HOTIs) has attracted increasing research interest~\cite{doi:10.1126/sciadv.aat0346, PhysRevLett.119.246401, PhysRevLett.122.256402, PhysRevLett.124.136407, Schindler2018, Yue2019, PhysRevX.8.031070, doi:10.1126/science.aah6442, PhysRevLett.119.246402, PhysRevLett.122.076801, PhysRevLett.127.196801, PhysRevLett.123.256402, PhysRevB.96.245115, Lee2020, PhysRevLett.120.026801, PhysRevB.97.155305, PhysRevB.97.205135, PhysRevB.97.205136, PhysRevX.9.011012, PhysRevLett.124.036803, PhysRevB.99.245151, PhysRevB.100.235302, PhysRevB.102.241102, PhysRevB.104.245302, PhysRevB.98.201114, PhysRevLett.124.166804, SerraGarcia2018, Noh2018, Peterson2018, Xue2019, Ni2019, Imhof2018, PhysRevLett.127.066801, PhysRevB.92.085126, arxiv.2109.00879}. In comparison with conventional first-order topological insulators, the HOTIs possess topologically localized states that are at least two dimensions lower than the bulk, namely, an $n$th-order topological insulator is characterized by $(d-n)$-dimensional gapless boundary states for $d$-dimensional bulk system. The HOTIs can be realized in a wide variety of systems, such as bismuth~\cite{Schindler2018}, phononic~\cite{SerraGarcia2018}, photonic~\cite{Noh2018, arxiv.2109.00879}, microwave~\cite{Peterson2018}, acoustic metamaterial~\cite{Xue2019, Ni2019}, and electrical circuit systems~\cite{Imhof2018}. In two-dimensional (2D) geometry, the HOTIs refer to second-order topological insulators. A second-order topological insulator in 2D geometry has a gapped 2D bulk state and gapped one-dimensional (1D) boundary state, but has topologically protected gapless zero-dimensional (0D) corner states localized at the intersections between 1D boundaries~\cite{doi:10.1126/science.aah6442, PhysRevLett.119.246402, PhysRevLett.123.256402, PhysRevB.96.245115, Lee2020, PhysRevLett.120.026801, PhysRevB.97.155305, PhysRevB.97.205135, PhysRevB.97.205136, PhysRevX.9.011012, PhysRevLett.124.036803, PhysRevB.99.245151, PhysRevB.100.235302, PhysRevB.102.241102, PhysRevB.104.245302, PhysRevB.98.201114, PhysRevLett.124.166804, SerraGarcia2018, Noh2018, Peterson2018, Xue2019, Ni2019, Imhof2018}.

In addition to Euclidean geometry, non-Euclidean geometry also possesses abundant novel physical phenomena. Hyperbolic lattices with constant negative curvature are promising platforms for fault-tolerant quantum computing~\cite{Pastawski2015, 7456305, Breuckmann2017, Lavasani2019universallogical, Jahn2021}. Inspired by recent experimental realizations of hyperbolic lattices in circuit quantum electrodynamics~\cite{Kollar2019}, a considerable number of theoretical studies and experimental observations of hyperbolic lattices have been reported~\cite{doi:10.1126/sciadv.abe9170, Ikeda2021, doi:10.1073/pnas.2116869119, PhysRevLett.129.088002, PhysRevB.105.125118, PhysRevE.106.034114,PhysRevLett.125.053901, Zhang2022, PhysRevB.105.245301, PhysRevLett.129.246402, 10.1038/s41467-023-36359-6, RUZZENE2021101491, PhysRevB.106.155146, PhysRevLett.128.166402, PhysRevB.106.155120, Lenggenhager2022,Kollar2020CMP, arxiv.2108.08854,PhysRevLett.128.013601,Zhu2021, PhysRevA.102.032208}. Based on the Riemann surface theory and algebraic geometry framework, a hyperbolic lattice generalization of Bloch band theory and crystallography has been proposed~\cite{doi:10.1126/sciadv.abe9170, Ikeda2021, doi:10.1073/pnas.2116869119, PhysRevB.105.125118, PhysRevE.106.034114, PhysRevLett.129.088002}. Recently, the non-Euclidean generalization of topological phenomena has also received enthusiastic attention~\cite{PhysRevLett.125.053901, Zhang2022, PhysRevB.105.245301, PhysRevLett.129.246402, 10.1038/s41467-023-36359-6, RUZZENE2021101491, PhysRevB.106.155146,PhysRevLett.128.166402}. Yu \emph{et~al}. proposed a non-Euclidean analog of the quantum spin Hall effect in hyperbolic lattices~\cite{PhysRevLett.125.053901}. Subsequently, the topological phase transition and transport properties of Chern insulator in a hyperbolic lattice were explored~\cite{PhysRevB.105.245301}. Urwyler \emph{et~al}. investigated the topological aspects arising in hyperbolic band theory based on the hyperbolic Haldane model and the hyperbolic Kane-Mele model~\cite{PhysRevLett.129.246402}. It is noted that an experimental demonstration of topological hyperbolic lattices on an electric circuit network has been reported by Zhang \emph{et~al}.~\cite{Zhang2022}. In their work, in addition to the theoretical proposal and experimental simulation of the hyperbolic Haldane model, they also showed that the fractal-like midgap higher-order zero-energy modes appear in deformed hyperbolic lattices, in which a pair of distinct coupling strengths is introduced~\cite{Zhang2022}. Motivated by the recent theoretical progress and experimental realization in topological hyperbolic lattices, further exploring the non-Euclidean generalization of higher-order topological phenomena is a fascinating issue. In this paper, we aim to investigate a non-Euclidean analog of HOTIs. We construct a HOTI model in hyperbolic lattices by introducing a time-reversal symmetry (TRS) breaking Wilson mass term into the quantum spin Hall insulator. For concreteness, the hyperbolic $\{4,5\}$, $\{8,3\}$, and $\{12,3\}$ lattices are considered, respectively. The non-Euclidean higher-order topological behavior is characterized by zero-energy effective corner states appearing in hyperbolic lattices. We use a real-space quadrupole moment to characterize the higher-order topology of the hyperbolic lattice with four zero-energy effective corner states. The symmetry analysis and robustness against weak disorder further confirm the topological nature of hyperbolic zero-energy effective corner states. It is noted that the hyperbolic system proposed in this paper breaks the translation symmetry, therefore the crystal momentum is no longer a good quantum number. On the other hand, we find that the value of the edge gap in the hyperbolic system decreases with increasing system size, which implies that the edge gap may vanish in the thermodynamic limit (see Appendix~\ref{AppendixFSE} for more details). These issues will be investigated in our future work.

The rest of the paper is organized as follows. In Sec.~\ref{HOTI}, we introduce a HOTI model in real space into hyperbolic $\{4, 5\}$, $\{8, 3\}$, and $\{12, 3\}$ lattices, and calculate the energy spectrum of the system and the probability distribution of the (near) zero-energy eigenstates when the Wilson mass term is turned off or on. In Sec.~\ref{Qxy}, we introduce the numerical calculations of the real-space quadrupole moment. In Sec.~\ref{perturbation}, we analyze the symmetry of hyperbolic HOTI systems with different types of perturbations, and show the energy spectrum under the effect of different perturbations. In Sec.~\ref{disorder}, we demonstrate the robustness of zero-energy effective corner states against symmetry-preserving weak disorder. Finally, we summarize our conclusions in Sec.~\ref{Conclusion}.

\section{Zero-energy effective corner states in hyperbolic lattices}
\label{HOTI}

In this paper, we adopt Schl\"{a}fli symbol $\{p, q\}$ to label the hyperbolic lattice in the Poincar\'{e} disk, which is obtained by projecting a hyperbolic plane with constant negative curvature into the unit circle in Euclidean plane~\cite{coxeter1973regular, weeks2001shape, greenberg1993euclidean}. $p$ represents the number of vertices of regular polygons that are uniformly tessellated in hyperbolic space, and $q$ means the number of polygons adjacent to the same vertex. In Euclidean geometry, a regular polygon uniformly tessellated in a plane must satisfy the relation $(p-2)(q-2)=4$. This means that only three regular polygons $\{3, 6\}$, $\{4, 4\}$ and $\{6, 3\}$ can be uniformly tessellated in the Euclidean plane. However, there are infinite kinds of regular polygons in the hyperbolic plane, which is guaranteed by the relation $(p-2)(q-2)>4$. It provides support for the unique physical phenomena in hyperbolic lattices.

Quantum spin Hall insulators are a promising platform for realizing HOTIs~\cite{doi:10.1126/sciadv.aat0346, PhysRevLett.119.246401, PhysRevLett.124.036803, PhysRevB.102.241102, PhysRevB.104.245302, PhysRevLett.124.166804}. First, we introduce Bernevig-Hughes-Zhang (BHZ) model ~\cite{doi:10.1126/science.1148047, doi:10.1126/science.1133734} of quantum spin Hall insulators to hyperbolic lattice in the real space, which can be described by the following tight-binding model Hamiltonian~\cite{PhysRevLett.124.036803, PhysRevResearch.2.012067}:
\begin{align}
H_{0}=-\frac{1}{2}\sum_{\left <j,k \right >}c_{j}^{\dagger}it_{1}\left [s_{z}\tau_{x}\cos(\theta_{jk})+s_{0}\tau_{y}\sin(\theta_{jk})\right ]c_{k}\nonumber \\
-\frac{1}{2}\sum_{\left <j,k \right >}c_{j}^{\dagger}t_{2}s_{0}\tau_{z}c_{k}+\sum_{j}(M+2t_{2})c_{j}^{\dagger}s_{0}\tau_{z}c_{j},
\label{eq1}
\end{align}
where $c_{j}^{\dagger}$ and $c_{j}$ are the creation and annihilation operators of electrons on site $j$. $\theta_{jk}$ represents the polar angle of the vector from site $k$ to site $j$ in the Poincar\'{e} disk. $s_{0}$ is the identity matrix, $s_{x,y,z}$ and $\tau_{x,y,z}$ are the Pauli matrices representing spin and orbital, respectively. $M$ denotes the Dirac mass, $t_{1}$ is the spin-orbit coupling strength, and $t_{2}$ is the hopping amplitude. It is worth noting that Hamiltonian $H_{0}$ does not possess translation symmetry in hyperbolic lattices in the presence of the polar angles $\theta_{jk}$, because $\theta_{jk}$ are defined in the Poincar\'{e} disk in the Euclidean plane, not projected from hyperbolic plane ~\cite{PhysRevB.105.245301}. Therefore, we will apply the numerical calculation approach in the real space to explore the topological properties of the hyperbolic lattices.

\begin{figure}[t]
	\includegraphics[width=8.5cm]{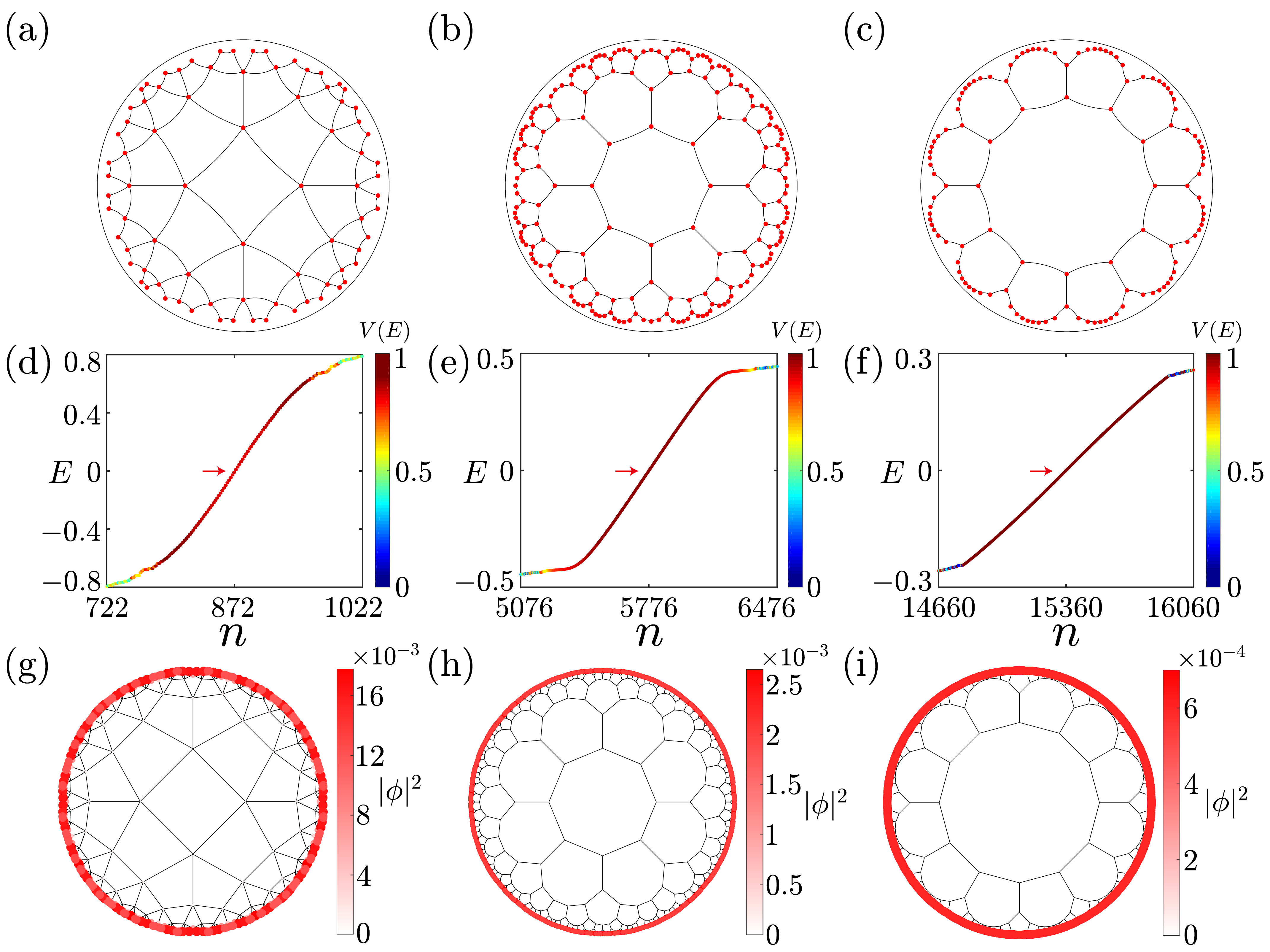} \caption{Schematic illustration of the hyperbolic (a) $\{4, 5\}$, (b) $\{8, 3\}$, and (c) $\{12, 3\}$ lattice. In the Poincar\'{e} disk, the solid black line means the distance between the nearest neighbors of two lattice sites, and any solid black line in the same disk represents the same distance. Energy spectrum of Hamiltonian $H_{0}$ on the hyperbolic (d) $\{4, 5\}$, (e) $\{8, 3\}$, and (f) $\{12, 3\}$ lattice. The color bar corresponds to the quantity $V(E)$ for the localization degree at the edge. (g) The probability distribution of four eigenstates near zero energy ($E_{n}=-0.0091,-0.0091,0.0091,0.0091$) marked by the red arrow in (d). The parameters are $M=-1$ and the lattice site number $N=436$ in (d) and (g). (h) The probability distribution of four eigenstates near zero energy ($E_{n}=-0.0012,-0.0012,0.0012,0.0012$) marked by the red arrow in (e). The parameters are $M=-1.2$ and $N=2888$ in (e) and (h). (i) The probability distribution of four eigenstates near zero energy ($E_{n}=-0.0005,-0.0005,0.0005,0.0005$) marked by the red arrow in (f). The parameters are $M=-1$ and $N=7680$ in (f) and (i). Here, we take the parameters $t_{1}=t_{2}=1$.}%
	\label{fig1}
\end{figure}

Before constructing higher-order topological states, we discuss the symmetries of Hamiltonian $H_{0}$. Similar to that in the Euclidean geometric system, Hamiltonian $H_{0}$ in the hyperbolic lattice possesses TRS $TH_{0}T^{-1}=H_{0}$ with TRS operator $T=is_{y}\tau_{0}\mathcal{K}$ (where $\mathcal{K}$ is the complex conjugation), particle-hole symmetry (PHS) $PH_{0}P^{-1}=-H_{0}$ with PHS operator $P=s_{z}\tau_{x}\mathcal{K}$, and chiral symmetry $SH_{0}S^{-1}=-H_{0}$ with chiral symmetry operator $S=PT$. In addition to the fundamental symmetries, Hamiltonian $H_{0}$ in the hyperbolic lattice also satisfies some spatial symmetries. For the hyperbolic $\{p, q\}$ lattice, $H_{0}$ has mirror symmetry $\left [ H_{0}, m_{z}\right ]=0$ with mirror symmetry operator $m_{z}=s_{z}\tau_{0}$ and $p$-fold rotational symmetry $\left [ H_{0}, C_{p}\right ]=0$. The $p$-fold rotational symmetry operator is $C_{p}=e^{-i\frac{\pi}{p}s_{z}\tau_{z}}\mathcal{R}_{p}$, where $\mathcal{R}_{p}$ is an orthogonal matrix permuting the vertices of the hyperbolic $\{p, q\}$ lattice to rotate the whole system by an angle of $2\pi/p$.

Here, for concreteness, the hyperbolic $\{4,5\}$, $\{8,3\}$, and $\{12,3\}$ lattices are constructed in the Poincar\'{e} disk as shown in Figs.~\ref{fig1}(a)--\ref{fig1}(c), respectively. Energy spectra of Hamiltonian $H_{0}$ are plotted in Figs.~\ref{fig1}(d)--\ref{fig1}(f). To quantify the localization degree of each energy on the edge, a quantity $V(E)=\sum_{j \in N_{\rm{edge}}}|\phi_{j}(E)|^{2}/\sum_{j \in N_{\rm{total}}}|\phi_{j}(E)|^{2}$ of energy $E$ is calculated~\cite{Zhang2022}, where $\phi_{j}(E)$ is the element of the eigenstate with the energy $E$ at site $j$, $N_{\rm{total}}$ represents all the sites of the sample, and $N_{\rm{edge}}$ refers to the sites that locate at the edge. In Figs.~\ref{fig1}(d)--\ref{fig1}(f), there are $248$, $2120$, and $6708$ sites at the edge, respectively. The color bars in Figs.~\ref{fig1}(d)--\ref{fig1}(f) represent the quantity $V(E)$. In previous work, the gapless chiral edge states were presented when the Chern insulator model was applied to a hyperbolic lattice ~\cite{PhysRevB.105.245301}. Here, we plot the probability distribution of the edge states of the BHZ model $H_{0}$ in the hyperbolic $\{4, 5\}$, $\{8, 3\}$, and $\{12, 3\}$ lattices, as shown in Figs.~\ref{fig1}(g)--\ref{fig1}(i). It is found that gapless edge states exist in three different hyperbolic lattices, and the double degeneracy of the edge states is guaranteed by TRS.

One of the ways to realize HOTIs is to construct mass domain walls on the boundaries of first-order topological insulators~\cite{PhysRevLett.124.036803, PhysRevLett.119.246402, PhysRevX.9.011012, doi:10.1126/sciadv.aat0346, PhysRevB.97.205136, PhysRevLett.119.246401, PhysRevB.97.205135, PhysRevX.8.031070, PhysRevB.102.241102, PhysRevB.104.245302}. The energy gap of the first-order topological insulator can be opened by breaking the TRS, therefore we construct the following Hamiltonian ~\cite{PhysRevLett.124.036803, PhysRevResearch.2.012067}:
\begin{align}
H_{1}&=H_{0}+H_{g},
\label{eq2}
\end{align}
with
\begin{align}
H_{g}&=\frac{g}{2}\sum_{\left <j,k \right >}c_{j}^{\dagger}\cos(\eta \theta_{jk})s_{x}\tau_{x}c_{k},
\label{eq2}
\end{align}
where $H_{g}$ is a Wilson mass term and $g$ is the magnitude of the Wilson mass. $\eta$ can only take even numbers, which is used to adjust the variation period of the Wilson mass. It is worth noting that Hamiltonian $H_{1}$ becomes non-Hermitian when $\eta$ takes odd numbers. The TRS breaking term $H_{g}$ gaps out 1D edge states in the first-order topological insulator. In addition, since $\cos(\eta \theta_{jk})$ in $H_{g}$ is alternating positive and negative, $H_{g}$ can construct new mass domain walls on the edge. We give more details of Hamiltonian $H_{1}$ in Appendix~\ref{AppendixH1}.

\begin{figure}[t]
	\includegraphics[width=8cm]{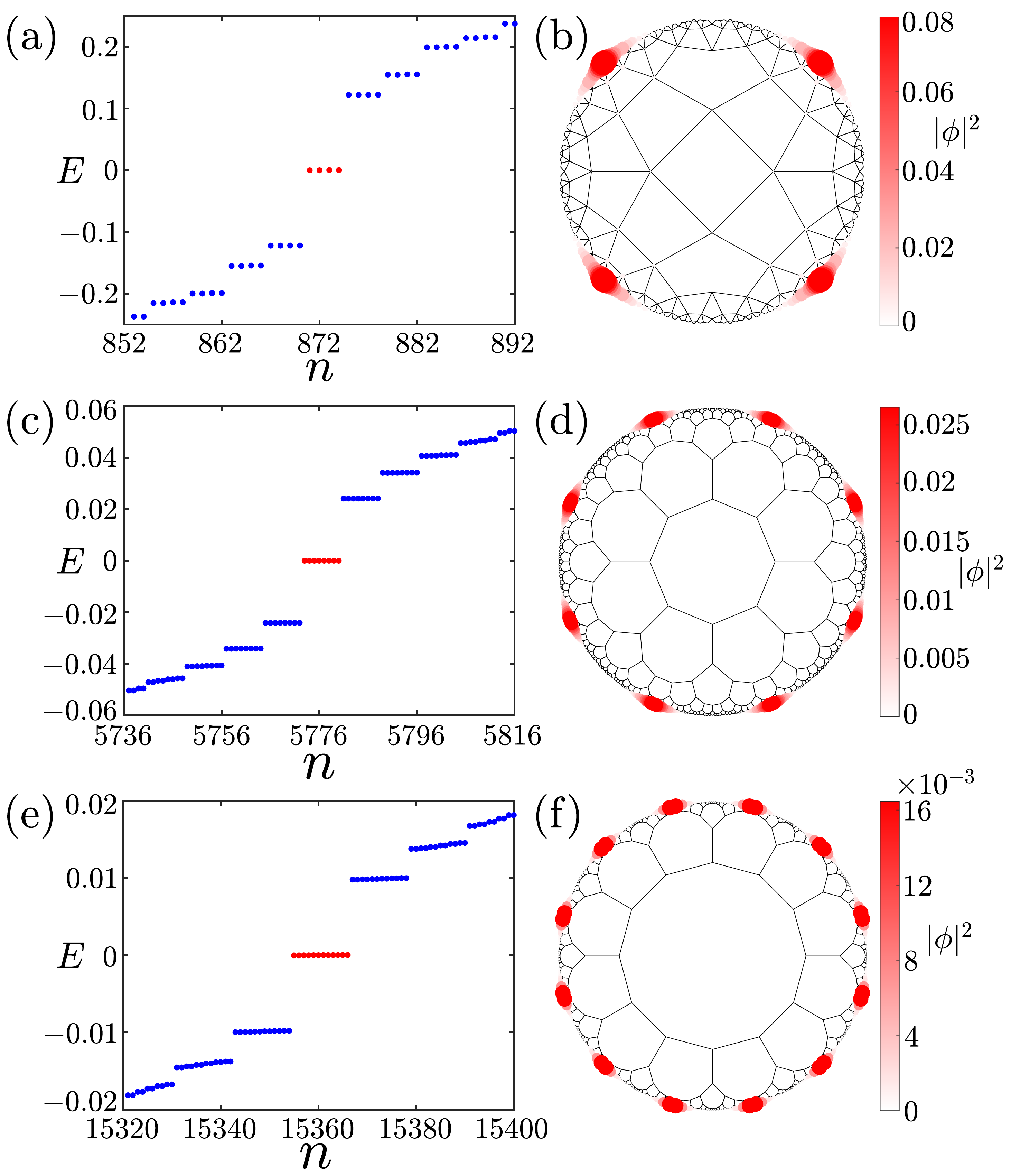} \caption{(a) Energy spectrum of Hamiltonian $H_{1}$ in the hyperbolic $\{4,5\}$ lattice. (b) The probability distribution of the 4 zero-energy eigenstates marked with red dots in (a). The parameters are $M=-1$, $g=0.5$, $\eta=2$, and $N=436$ in (a) and (b). (c) Energy spectrum of Hamiltonian $H_{1}$ in the hyperbolic $\{8,3\}$ lattice. (d) The probability distribution of the 8 zero-energy eigenstates marked with red dots in (c). The parameters are $M=-1.2$, $g=0.5$, $\eta=4$, and $N=2888$ in (c) and (d). (e) Energy spectrum of Hamiltonian $H_{1}$ in the hyperbolic $\{12,3\}$ lattice. (f) The probability distribution of the 12 zero-energy eigenstates marked with red dots in (e). The parameters are $M=-1$, $g=0.5$, $\eta=6$, and $N=7680$ in (e) and (f). Here, we take the parameters $t_{1}=t_{2}=1$.}%
	\label{fig2}
\end{figure}

Next, we turn on the $H_{g}$ term to explore higher-order topological states. Here we first set $\eta=2$ and use the Jackiw-Rebbi mechanism to analyze the newly generated Dirac-mass domain walls~\cite{PhysRevD.13.3398}. The sign change of $H_{g}$ is modulated by $\cos(\eta \theta_{jk})$, and when $\eta=2$, $H_{g}$ is divided into four parts with alternating positive and negative signs in the whole plane by $\cos(\eta \theta_{jk})$. Then, the boundary between the positive and negative regions of Dirac mass is the domain wall with the Dirac mass equal to zero. When $\eta=2$, there are four such domain walls. As shown in Fig.~\ref{fig2}(a), there are four zero energies marked with red dots in the energy gap opened by the $H_{g}$ term. In Fig.~\ref{fig2}(b), we show the probability distribution of these four zero-energy eigenstates and find that they are localized exactly at the new domain walls constructed by $H_{g}$. This is consistent with our analysis results. These four zero-energy modes are called hyperbolic effective corner states, since there are not real geometric corners (intersections between 1D boundaries) in the Poincar\'{e} disk corresponding to a hyperbolic lattice~\cite{Zhang2022}. Then, we apply Hamiltonian $H_{1}$ with $\eta=4$ and $\eta=6$ in the hyperbolic $\{8,3\}$ and $\{12,3\}$ lattices, respectively. By diagonalizing Hamiltonian $H_{1}$ in two different hyperbolic lattices, we obtain partial energies near zero energy, as shown in Figs.~\ref{fig2}(c) and \ref{fig2}(e). It can be found that there are 8 and 12 in-gap zero energies, respectively. Similarly, we calculate the probability distributions of these 8 (12) zero-energy eigenstates and find that they all exhibit a $0$D form localized at the edge of the hyperbolic lattice. The 8 and 12 zero-energy effective corner states in the hyperbolic $\{8,3\}$ and $\{12,3\}$ lattices are plotted in Figs.~\ref{fig2}(d) and \ref{fig2}(f), respectively. Numerical calculations show that the number of zero-energy effective corner states of a hyperbolic lattice can be tuned by changing the value of $\eta$. In hyperbolic systems, $\eta=p/2$ is not a necessary condition to produce the zero-energy effective corner states. There can also exit 12 zero-energy effective corner states in the hyperbolic $\{8,3\}$ lattice when the value of $\eta$ in Hamiltonian $H_{1}$ is taken as $\eta=6$ for the hyperbolic $\{8,3\}$ lattice. In Appendix~\ref{AppendixZEECS}, we discuss the existence of zero-energy effective corner states in the hyperbolic lattice with $\eta\neq p/2$, and compare the difference of symmetry between the two cases (i.e., $\eta=p/2$ and $\eta\neq p/2$).

\section{The Quadrupole moment}
\label{Qxy}
In Euclidean space, the quadrupole moment can be used as a topological invariant to characterize the higher-order topology, and it can be expressed in momentum space~\cite{doi:10.1126/science.aah6442, PhysRevB.96.245115} and real space~\cite{PhysRevLett.125.166801, PhysRevB.100.245134, PhysRevB.100.245135, PhysRevB.101.195309, PhysRevResearch.2.012067, PhysRevB.104.245302, PhysRevB.103.085408}, respectively. Due to the fact that the translation symmetry is lacking in the hyperbolic lattice model considered, we utilize the real-space quadrupole moment to characterize the topological properties of the zero-energy effective corner states. The real-space quadrupole moment $Q_{xy}$ is given by~\cite{PhysRevLett.125.166801, PhysRevB.100.245134, PhysRevB.100.245135, PhysRevB.101.195309, PhysRevResearch.2.012067, PhysRevB.104.245302, PhysRevB.103.085408}
\begin{align}
Q_{xy}&=\left [ \frac{1}{2\pi}{\rm{Im}}~{\rm{log}}~{\rm{det}}(\Psi_{\rm occ}^{\dagger}\hat{U}\Psi_{\rm occ})-Q_{0} \right ]~{\rm{mod}}~1,
\label{eqQxy}
\end{align}
with
\begin{align}
Q_{0}&=\frac{1}{2}\sum_{j}x_{j}y_{j}/A,
\label{eqQxy}
\end{align}
where $\Psi_{\rm occ}$ are the occupied eigenstates of $H_{1}$, $\hat{U}$ is a diagonal matrix whose diagonal elements are $e^{2\pi i x_{j}y_{j}/A}$, and $(x_{j},y_{j})$ denotes the rescaled coordinate of the $j$th site in the Poincar\'{e} disk. $A=\pi r^{2}$ is the area of the disk with the radius $r=1$. It is worth noting that the area $A$ is computed with respect to the Euclidean metric. In calculations, we need to translate the coordinates in interval $x_{j},y_{j}\in(-1,1)$ to interval $x_{j},y_{j}\in(0,2)$. A HOTI is characterized by the quadrupole moment $Q_{xy}=0.5$, while the quadrupole moment of a trivial system is equal to $0$.

\begin{figure}[t]
	\includegraphics[width=8.5cm]{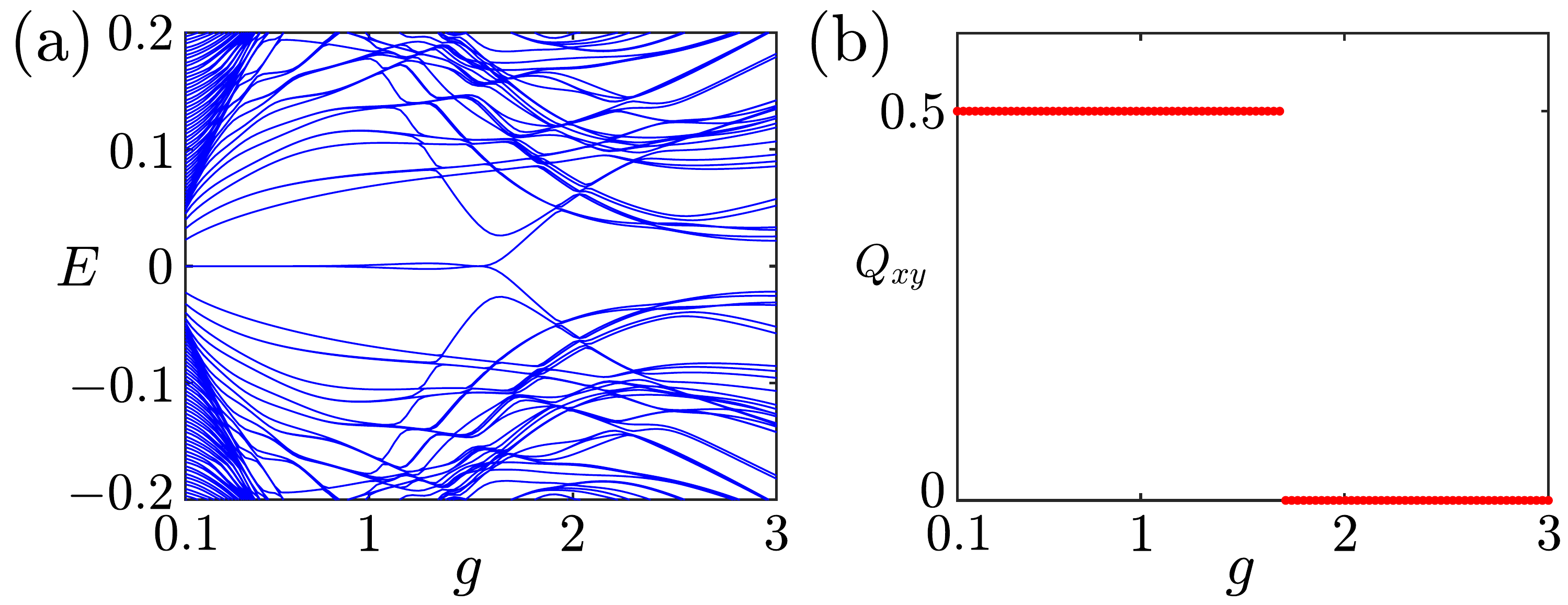} \caption{(a) The energy spectrum of Hamiltonian $H_{1}$ in the hyperbolic $\{4,5\}$ lattice with respect to $g$. (b) The quadrupole moment $Q_{xy}$ as a function of $g$. Here, we take the parameters $M=-1$, $t_{1}=t_{2}=1$, $\eta=2$, and $N=2320$.}%
	\label{fig3}
\end{figure}

In Fig.~\ref{fig3}(a), we show the evolution of the energy spectrum of Hamiltonian $H_{1}$ in the hyperbolic $\{4,5\}$ lattice with respect to the Wilson mass $g$. When the Wilson mass term $H_{g}$ is turned on, the system possesses four effective corner states at zero energy. While the strength of the Wilson mass term $H_{g}$ exceeds a certain value $g=1.565$, the zero-energy effective corner states are destroyed and an energy gap of the system appears. To explore this topological phase transition process, we calculate the real-space quadrupole moment as a function of Wilson mass $g$ as shown in Fig.~\ref{fig3}(b). It is found that the real-space quadrupole moment changes from $Q_{xy}=0.5$ to $Q_{xy}=0$ at $g=1.711$. It is obvious that the interval in which the zero-energy effective corner states exist can match well with the plateau of the quadrupole moment values with $Q_{xy}=0.5$. A slight difference of the phase transition point between Figs.~\ref{fig3}(a) and \ref{fig3}(b) is owing to the finite-size effect. Thus, the four zero-energy effective corner states, which are characterized by the real-space quadrupole moment, are a hallmark feature of the hyperbolic higher-order topology.

However, the real-space quadrupole moment does not apply to hyperbolic lattices with 8 or 12 zero-energy effective corner states. It is an intriguing topic to explore real-space topological invariants that can characterize the hyperbolic higher-order topology with more than four corner states.

\section{Stability of zero-energy effective corner states}

In this section, we will explore the effects of perturbation and disorder on the zero-energy effective corner states. First, we examine whether zero-energy effective corner states in hyperbolic lattices are protected by specific symmetries through symmetry analysis and the responses of zero-energy effective corner states to different perturbations. Then, we present the robustness of the zero-energy modes in the hyperbolic lattices against weak disorder.

\subsection{Symmetry analysis}
\label{perturbation}

\begin{figure*}[t]
	\includegraphics[width=13cm]{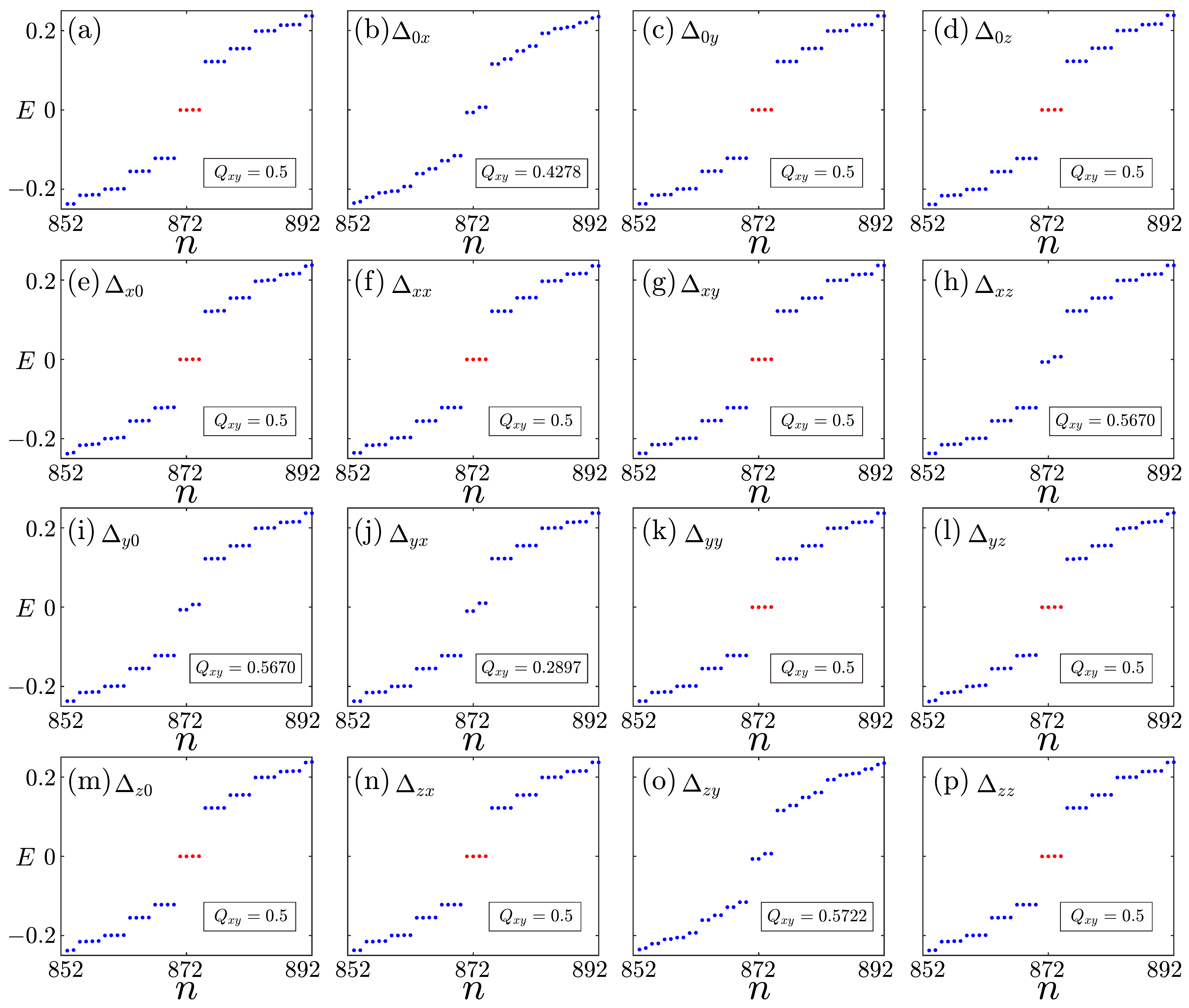} \caption{(a) Energy spectrum of Hamiltonian $H_{1}$ in the hyperbolic $\{4,5\}$ lattice. (b)--(p) Energy spectrum of $H_{1}+\Delta_{mn}$ in the hyperbolic $\{4,5\}$ lattice, where $m,n=0,x,y,z$. The quadrupole moment in different cases is shown in the insets. Here, we take the parameters $M=-1$, $t_{1}=t_{2}=1$, $g=0.5$, $\eta=2$, the perturbation strength $U=0.01$, and $N=436$.}%
	\label{fig4}
\end{figure*}

\begin{table}[b]
    \caption{Symmetry analysis of Hamiltonian $H_{1}$ without ($U=0$) and with ($U\neq0$) the perturbation terms $\Delta_{mn}$ on the hyperbolic $\{4,5\}$, $\{8,3\}$, and $\{12,3\}$ lattices. Here, for the hyperbolic $\{4,5\}$, $\{8,3\}$, and $\{12,3\}$ lattices, $C_{p}$ refers to $4$-fold, $8$-fold, and $12$-fold rotational symmetries, respectively. Check mark indicates that the symmetry in this case is preserved, and a cross mark means the symmetry is absent.}
\resizebox{\linewidth}{!}{
\begin{tabular}{ccccccccccccccccc}
\hline
\hline
\multirow{2}{*}{}&\multirow{2}{*}{$U=0$}&\multicolumn{15}{c}{$U\not=0$}\\
\cline{3-17}&&$0x$&$0y$&$0z$&$x0$&$xx$&$xy$&$xz$&$y0$&$yx$&$yy$&$yz$&$z0$&$zx$&$zy$&$zz$\\
\hline
$P$ &\checkmark      &$\times$     &$\times$      &\checkmark &\checkmark &\checkmark &\checkmark &$\times$ &$\times$      &$\times$      &$\times$     &\checkmark &$\times$       &$\times$      &$\times$      &\checkmark \\
\hline
$T$ &$\times$      &$\times$      &$\times$      &$\times$     &$\times$      &$\times$       &$\times$      &$\times$ &$\times$      &$\times$      &$\times$      &$\times$     &$\times$       &$\times$      &$\times$       &$\times$ \\
\hline
$S$ &$\times$      &$\times$      &$\times$      &$\times$      &$\times$      &$\times$      &$\times$      &$\times$ &$\times$      &$\times$      &$\times$      &$\times$      &$\times$      &$\times$      &$\times$      &$\times$ \\
\hline
$m_{x}$ &\checkmark &\checkmark &\checkmark &\checkmark &\checkmark &\checkmark &\checkmark &\checkmark &$\times$      &$\times$      &$\times$      &$\times$      &$\times$       &$\times$      &$\times$       &$\times$ \\
\hline
$m_{y}$ &\checkmark &$\times$      &$\times$      &\checkmark &$\times$       &\checkmark &\checkmark &$\times$ &\checkmark &$\times$      &$\times$      &\checkmark &$\times$       &\checkmark &\checkmark &$\times$ \\
\hline
$m_{z}$ &$\times$      &$\times$      &$\times$      &$\times$       &$\times$       &$\times$      &$\times$      &$\times$ &$\times$      &$\times$      &$\times$       &$\times$      &$\times$        &$\times$      &$\times$      &$\times$ \\
\hline
$Sm_{z}$ &$\checkmark$ &$\times$      &$\checkmark$       &$\checkmark$ &$\checkmark$       &$\checkmark$       &$\times$      &$\times$ &$\times$      &$\times$      &$\checkmark$       &$\checkmark$      &\checkmark   &$\checkmark$       &$\times$      &$\times$ \\
\hline
$C_{p}$ &$\times$      &$\times$      &$\times$      &$\times$       &$\times$       &$\times$      &$\times$      &$\times$ &$\times$      &$\times$      &$\times$      &$\times$       &$\times$        &$\times$      &$\times$     &$\times$ \\
\hline
$C_{p}T$ &\checkmark &$\times$      &$\times$      &\checkmark &$\times$        &$\times$      &\checkmark &$\times$ &$\times$      &$\times$      &\checkmark &$\times$      &$\times$        &$\times$      &$\times$       &$\times$ \\
\hline
$C_{p}m_{x}$ &$\times$      &$\times$       &$\times$      &$\times$      &$\times$        &$\times$      &$\times$      &$\times$ &$\times$      &$\times$       &$\times$      &$\times$      &$\times$        &$\times$       &$\times$      &$\times$ \\
\hline
$C_{p}m_{y}$ &$\times$      &$\times$       &$\times$      &$\times$      &$\times$       &$\times$       &$\times$      &$\times$ &$\times$      &$\times$       &$\times$      &$\times$       &$\times$       &$\times$       &$\times$     &$\times$ \\
\hline
$C_{p}m_{z}$ &\checkmark &$\times$      &$\times$       &\checkmark &$\times$       &$\times$       &$\times$      &$\times$ &$\times$      &$\times$      &$\times$       &$\times$      &\checkmark   &$\times$       &$\times$      &\checkmark \\
\hline
\hline
\end{tabular}
}
    \label{tab1}
\end{table}

It is known that, by adding perturbations to the system, the gapless topological boundary states remain stable as long as the symmetry protecting the topological system is preserved. In this subsection, we add a uniform perturbation $\Delta_{mn}=U\sum_{j}c_{j}^{\dagger}s_{m}\tau_{n}c_{j}$ to Hamiltonian $H_{1}$ to examine the stability of zero-energy effective corner states. $U$ is the strength of the perturbation, and $m,n=0,x,y,z$. Here, for concreteness, we take the $\{4,5\}$ lattice as an example to show the changes of the near-zero energy spectrum under the effect of different perturbations. The cases of the $\{8,3\}$ lattice and the $\{12,3\}$ lattice are similar to those in the $\{4,5\}$ lattice. In addition, we calculate the quadrupole moment of the system under the effect of different perturbations respectively. As shown in Fig.~\ref{fig4}, the zero-energy modes remain stable in the presence of ten kinds of uniform perturbations, which are $\Delta_{0y}$ [Fig.~\ref{fig4}(c)], $\Delta_{0z}$ [Fig.~\ref{fig4}(d)], $\Delta_{x0}$ [Fig.~\ref{fig4}(e)], $\Delta_{xx}$ [Fig.~\ref{fig4}(f)], $\Delta_{xy}$ [Fig.~\ref{fig4}(g)], $\Delta_{yy}$ [Fig.~\ref{fig4}(k)], $\Delta_{yz}$ [Fig.~\ref{fig4}(l)], $\Delta_{z0}$ [Fig.~\ref{fig4}(m)], $\Delta_{zx}$ [Fig.~\ref{fig4}(n)], and $\Delta_{zz}$ [Fig.~\ref{fig4}(p)], respectively. The other five kinds of perturbations can destroy the zero-energy modes. It is found that the quadrupole moment $Q_{xy}=0.5$ for the system in which the zero energy mode remains stable, while the quadrupole moment of the system in which the zero energy mode is destroyed deviates from the quantized value $0.5$. Like conventional topological insulators, topologically localized states in HOTIs are likewise protected by specific symmetries. Next, we explore the symmetries that protect zero-energy effective corner states in hyperbolic lattices by analyzing the changes in symmetry under perturbation.

\begin{table}[b]
    \caption{Symmetry analysis of the Hamiltonian $H_{1}$ with the combined perturbation terms $\Delta_{kl,mn}=\sum_{j}c_{j}^{\dagger}\left ( U_{1}s_{k}\tau_{l}+U_{2}s_{m}\tau_{n}\right ) c_{j}$ on the hyperbolic $\{4,5\}$, $\{8,3\}$, and $\{12,3\}$ lattices. Here, for the hyperbolic $\{4,5\}$, $\{8,3\}$, and $\{12,3\}$ lattices, $C_{p}$ refers to $4$-fold, $8$-fold, and $12$-fold rotational symmetries, respectively. Check mark indicates that the symmetry in this case is preserved, and a cross mark means the symmetry is absent.}
\begin{tabular}{ccccc}
\hline
\hline
&$\Delta_{yz,zz}$&$\Delta_{yy,z0}$&$\Delta_{xy,yy}$&$\Delta_{z0,zz}$\\
\hline
$P$ &\checkmark      &$\times$     &$\times$      &$\times$ \\
\hline
$Sm_{z}$ &$\times$ &\checkmark      &$\times$       &$\times$\\
\hline
$C_{p}T$ &$\times$ &$\times$      &\checkmark      &$\times$\\
\hline
$C_{p}m_{z}$ &$\times$ &$\times$      &$\times$       &\checkmark\\
\hline
\hline
\end{tabular}
    \label{tab2}
\end{table}

\begin{figure}[t]
	\includegraphics[width=8.5cm]{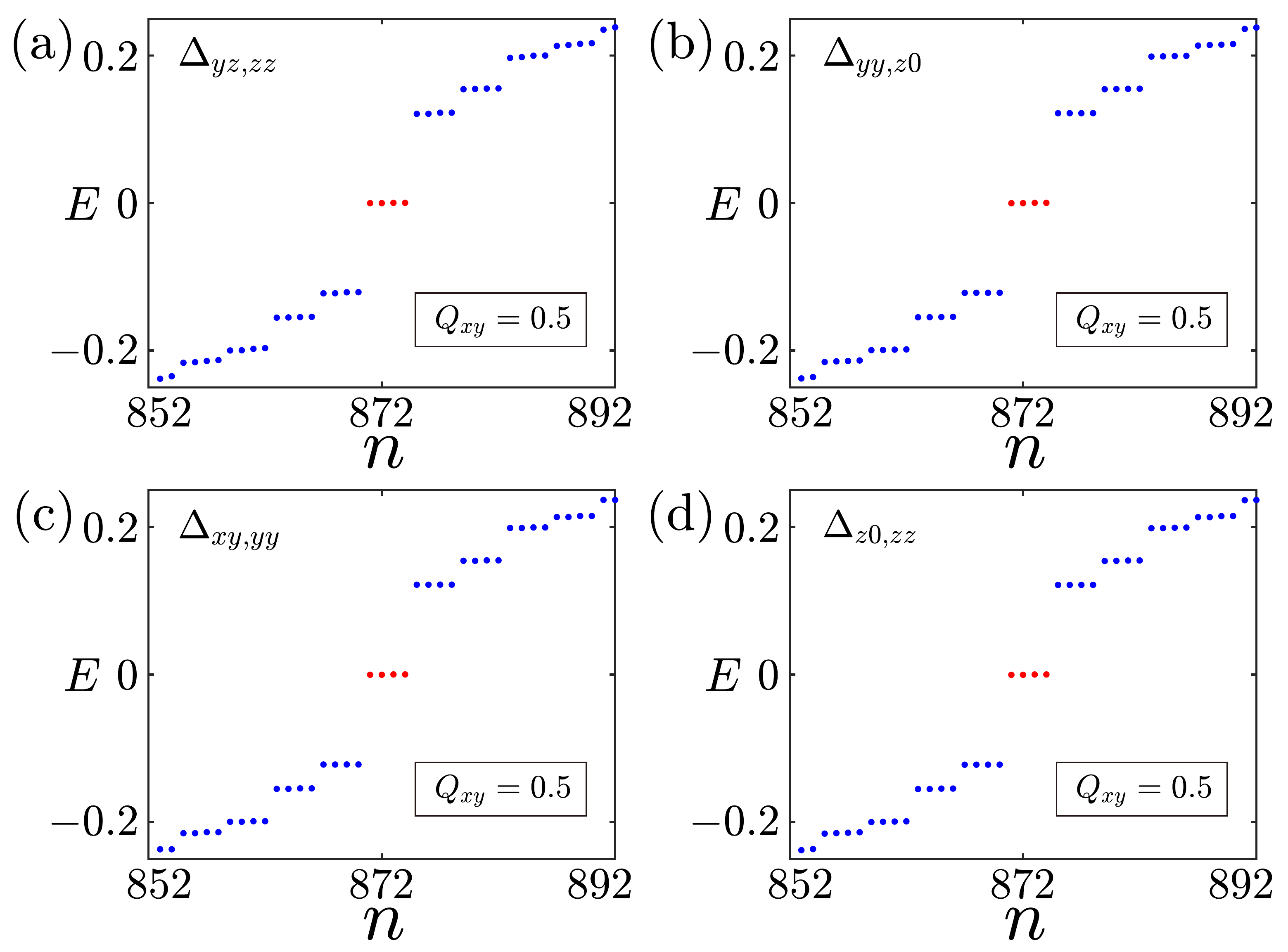} \caption{(a)--(d) Energy spectrum of $H_{1}+\Delta_{kl, mn}$ in the hyperbolic $\{4,5\}$ lattice, in which the type of combined perturbation is labeled. The quadrupole moment in different cases is shown in the insets. Here, we take the parameters $M=-1$, $t_{1}=t_{2}=1$, $g=0.5$, $\eta=2$, the perturbation strength $U_{1}=0.01$, $U_{2}=0.02$, and $N=436$.}%
	\label{fig5}
\end{figure}

The results of symmetry analysis are presented in Table~\ref{tab1}. In the absence of perturbation $\Delta_{mn}$, Hamiltonian $H_{1}$ satisfies PHS, mirror symmetries $m_{x}$ and $m_{y}$, and combined symmetries $Sm_{z}$, $C_{p}T$, $C_{p}m_{z}$. Comparing with Fig.~\ref{fig4}, we find that the zero-energy modes remain stable when one of the PHS $P$, the effective chiral symmetry $S_{\rm eff}=Sm_{z}$, and combined symmetries $C_{p}T$ and $C_{p}m_{z}$ is preserved, while the zero-energy modes are broken when these four symmetries are absent simultaneously. When perturbations is considered, the system respects at least two symmetries protecting the corner modes except for $\Delta_{0y}$ and $\Delta_{zx}$. Therefore, to verify whether each of these four symmetries can provide protection for the zero-energy states alone, we exclude the interference of redundant symmetries by combining two different perturbations. In Table~\ref{tab2}, we show the results of the symmetry analysis of Hamiltonian $H_{1}$ under the effect of four combined perturbations $\Delta_{kl,mn}=\sum_{j}c_{j}^{\dagger}\left ( U_{1}s_{k}\tau_{l}+U_{2}s_{m}\tau_{n}\right ) c_{j}$. When combined $\Delta_{yz,zz}$ perturbation is applied to Hamiltonian $H_{1}$, the system preserves only PHS $P$. Similarly, when $\Delta_{yy,z0}$, $\Delta_{xy,yy}$, and $\Delta_{z0,zz}$ perturbation are applied to Hamiltonian $H_{1}$, the system preserves only $Sm_{z}$, $C_{p}T$, and $C_{p}m_{z}$, respectively. The energy spectrum of Hamiltonian $H_{1}$ in the hyperbolic $\{4,5\}$ lattice under the effect of four combined perturbations is shown in Fig.~\ref{fig5}, and we calculate the quadrupole moment of these systems respectively. After eliminating redundant symmetries, we find that any one of these four symmetries preserves the topological properties of the zero-energy effective corner states. This is in agreement with the previously discussed results and furthermore shows that any one of the PHS $P$, effective chiral symmetry $S m_{z}$, $C_{p}T$, and $C_{p}m_{z}$ can protect the zero-energy effective corner states. Moreover, from the perspective of the quadrupole moment, these symmetries can guarantee that the quadrupole moment is quantized. When these four symmetries are broken simultaneously, the quadrupole moment deviates from the quantization ($0$ or $0.5$).

\subsection{Robustness of the zero-energy modes against disorder}
\label{disorder}

\begin{figure}[t]
	\includegraphics[width=8.5cm]{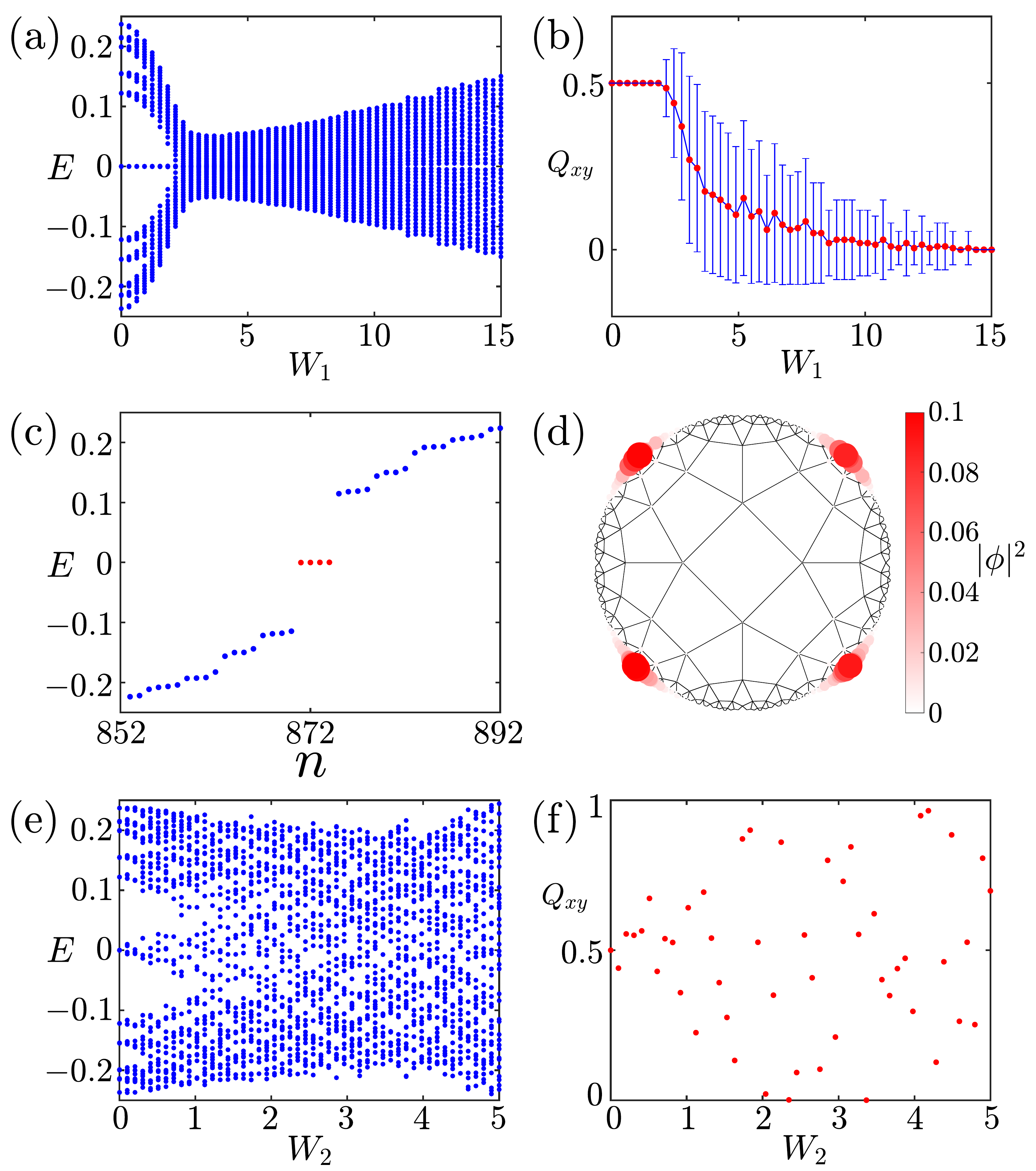} \caption{The $s_{0}\tau_{z}$-type disorder: (a) Eigenenergy of $H_{1}+\Delta H_{1}$ as a function of the disorder strength $W_1$ in the hyperbolic $\{4,5\}$ lattice. (b) The quadrupole moment $Q_{xy}$ as a function of $W_{1}$. The error bar represents the standard deviation of $100$ samples. (c) Energy spectrum in the hyperbolic $\{4,5\}$ lattice when $W_{1}=0.5$. (d) The probability distribution of the four zero-energy modes [the red dots in (c)]. The $s_{0}\tau_{x}$-type disorder: (e) Eigenenergy of $H_{1}+\Delta H_{2}$ as a function of the disorder strength $W_{2}$ in the hyperbolic $\{4,5\}$ lattice. (f) The quadrupole moment $Q_{xy}$ as a function of $W_{2}$. In (e) and (f), we take only one sample. Here, we take the parameters $M=-1$, $t_{1}=t_{2}=1$, $g=0.5$, $\eta=2$, and $N=436$.}%
	\label{fig6}
\end{figure}

Now, we study the zero-energy effective corner states in the hyperbolic lattices with disorder. Here, we take two kinds of on-site disorder that is randomly distributed in the whole sample as examples, which are, respectively, given by
\begin{align}
\Delta H_1&=W_1\sum_{j}c_{j}^{\dagger}\omega_{j}s_{0}\tau_{z}c_{j},\\
\Delta H_2&=W_2\sum_{j}c_{j}^{\dagger}\omega_{j}s_{0}\tau_{x}c_{j},
\label{eq2}
\end{align}
where $W_{1,2}$ depict the disorder strength, and $\omega_{j}$ is uniformly distributed within $[-0.5,0.5]$.
Both of them break the symmetries associated with the rotation symmetry including $C_{p}T$ and $C_{p}m_{z}$. The difference between $\Delta H_1$ and $\Delta H_2$ is that the former preserves the PHS $P$ and the effective chiral symmetry $S m_z$, while the latter breaks them.

In Fig.~\ref{fig6}(a), we show the evolution of the energy spectrum in the hyperbolic $\{4,5\}$ lattice for the disorder strength $W_1$ with $W_2=0$. The four degenerate zero-energy states remain stable in weak disorder until the disorder strength exceeds a certain value $W_1=1.837$. Accordingly, we calculate the quadrupole moment $Q_{xy}$ as a function of the disorder strength $W_{1}$ as shown in Fig.~\ref{fig6}(b). It is obvious that the quadrupole moment maintains a quantized plateau of $Q_{xy}=0.5$ when the zero-energy states remains stable. This phenomenon is observed more clearly by plotting the spectrum and the probability distribution of the zero-energy states for $W_1=0.5$ [Figs.~\ref{fig6}(c) and \ref{fig6}(d)].

However, this robustness is not valid for all types of disorders. For $W_1=0$ and $W_2\neq 0$, the weak $s_{0}\tau_{x}$-type disorder destroys the degenerate zero-energy states immediately, and the quadrupole moment deviates from the quantization [Figs.~\ref{fig6}(e) and \ref{fig6}(f)]. Therefore, we show that the zero-energy effective corner states are robust against weak disorder when $P$ and $S m_{z}$ are preserved.

\section{Conclusion}
\label{Conclusion}

In this paper, we investigate higher-order topological behavior of the hyperbolic lattices by introducing a TRS breaking Wilson mass term into the quantum spin Hall insulator. Three hyperbolic $\{4,5\}$, $\{8,3\}$, and $\{12,3\}$ lattices are considered, respectively. By adjusting the variation period of the TRS breaking term, different numbers zero-energy effective corner states can appear in the hyperbolic lattices. Based on numerical calculations of real-space quadrupole moment in hyperbolic $\{4,5\}$ lattices, the system with these zero-energy states are demonstrated to be HOTIs. Although the real-space quadrupole moment can only characterize the topological properties of the hyperbolic lattice with four zero-energy effective corner states, we believe that the zero-energy states in other hyperbolic lattices with similar responses to perturbation and disorder are also topological. However, we do not construct a topological invariant that can characterize higher-order topological properties of hyperbolic lattices with more zero-energy states, and we will continue to investigate them in future work.

By analyzing the effects of different types of perturbations on the system, we find that the hyperbolic zero-energy effective corner states are protected by four symmetries, which are PHS $P$, effective chiral symmetry $S_{\rm eff}$, and combined symmetries $C_{p}T$ and $C_{p}m_{z}$. Moreover, we find that these zero-energy states are robust against the weak disorder that preserves these four symmetries, but they are broken under the effect of the disorder that breaks these four symmetries. Although our model does not possess translational symmetry in the projected hyperbolic lattice, we demonstrate that the zero-energy effective corner states in the hyperbolic lattice are topologically nontrivial using a numerical calculation approach in the real space. We will continue to study hyperbolic HOTIs with translation symmetry in our future work.

\emph{Note added.} Recently, we became aware of a complementary study~\cite{arxiv.2209.02262}, which addresses similar problems from a different perspective.

\section*{Acknowledgments}
B.Z. was supported by the NSFC (Grant No. 12074107), the program of outstanding young and middle-aged scientific and technological innovation team of colleges and universities in Hubei Province (Grant No. T2020001) and the innovation group project of the Natural Science Foundation of Hubei Province of China (Grant No. 2022CFA012). C.-B.H was supported by the Doctoral Research Start-Up Fund of Hubei University of Science and Technology (Grant No. BK202316). T.P. was supported by the Doctoral Research Start-Up Fund of Hubei University of Automotive Technology (Grant No. BK202216).

\appendix

\section{Construction of the Hamiltonian $H_{1}$}
\label{AppendixH1}
In this appendix, we present the details of Hamiltonian $H_{1}$. Before presenting the construction details of $H_{1}$, let us review some concepts in the Poincar\'{e} disk model. The sites in the Poincar\'{e} disk can be considered as all points of a unit disk $\mathbb{D}=\{ z\in \mathbb{C},~|z|<1 \}$. Therefore, the sites in the disk can be expressed as $z=x+iy=re^{i\theta}$. The hyperbolic distance between two points $z,z'\in \mathbb{D}$ is given by~\cite{PhysRevB.105.125118}
\begin{align}
d(z,z')=\kappa~{\rm{arcosh}}\left ( 1+\frac{2|z-z'|^{2}}{(1-|z|^{2})(1-|z'|^{2})} \right ),
\end{align}
where $\kappa$ is the curvature radius. Next, we discuss the details in Hamiltonian $H_{1}$. Hamiltonian $H_{1}$ is described as the following equation:
\begin{align}
H_{1}=&-\frac{1}{2}\sum_{\left <j,k \right >}c_{j}^{\dagger}it_{1}\left [s_{z}\tau_{x}\cos(\theta_{jk})+s_{0}\tau_{y}\sin(\theta_{jk})\right ]c_{k}\nonumber \\
&-\frac{1}{2}\sum_{\left <j,k \right >}c_{j}^{\dagger}t_{2}s_{0}\tau_{z}c_{k}+\sum_{j}(M+2t_{2})c_{j}^{\dagger}s_{0}\tau_{z}c_{j}\nonumber \\
&+\frac{g}{2}\sum_{\left <j,k \right >}c_{j}^{\dagger}\cos(\eta \theta_{jk})s_{x}\tau_{x}c_{k},
\end{align}
where $c_{j}^{\dagger}$ and $c_{j}$ are the creation and annihilation operators of electrons on site $j$. $\theta_{jk}$ represents the polar angle of the vector from site $k$ to site $j$ in the Poincar\'{e} disk, which is given by $z_{j}-z_{k}=r_{jk}e^{i\theta_{jk}}$. $s_{0}$ is the identity matrix, $s_{x,y,z}$ and $\tau_{x,y,z}$ are the Pauli matrices representing spin and orbital, respectively. $M$ denotes the Dirac mass, $t_{1}$ is the spin-orbit coupling strength, and $t_{2}$ is the hopping amplitude between the nearest neighbor sites. The nearest neighbor distance here refers to the nearest hyperbolic distance between two sites in the Poincar\'{e} disk. The last term $H_{g}=\frac{g}{2}\sum_{\left <j,k \right >}c_{j}^{\dagger}\cos(\eta \theta_{jk})s_{x}\tau_{x}c_{k}$ is the Wilson mass term, where $\eta$ is used to tune the variation period of the Wilson mass. $\eta$ can only take even numbers. If a Hermitian system satisfies $H_{g}=H_{g}^{\dagger}$, then $\cos(\eta \theta_{jk})=\cos(\eta \theta_{kj})=\cos[\eta (\theta_{jk}+\pi)]$. When $\eta$ takes an odd number, this equation is not satisfied and the Hamiltonian becomes non-Hermitian.

\section{Zero-energy effective corner states with $\eta\neq p/2$}
\label{AppendixZEECS}
In the main text, we discuss the case where $\eta=p/2$. In fact, when $\eta$ is not equal to $p/2$, the zero-energy effective corner states can also occur in the hyperbolic lattices. In this appendix, we discuss the electronic structure of the system when $\eta\neq p/2$. Moreover, by symmetry analysis, we compare the difference of symmetry of the systems with $\eta=p/2$ and $\eta\neq p/2$.

\begin{figure}[t]
	\includegraphics[width=8.5cm]{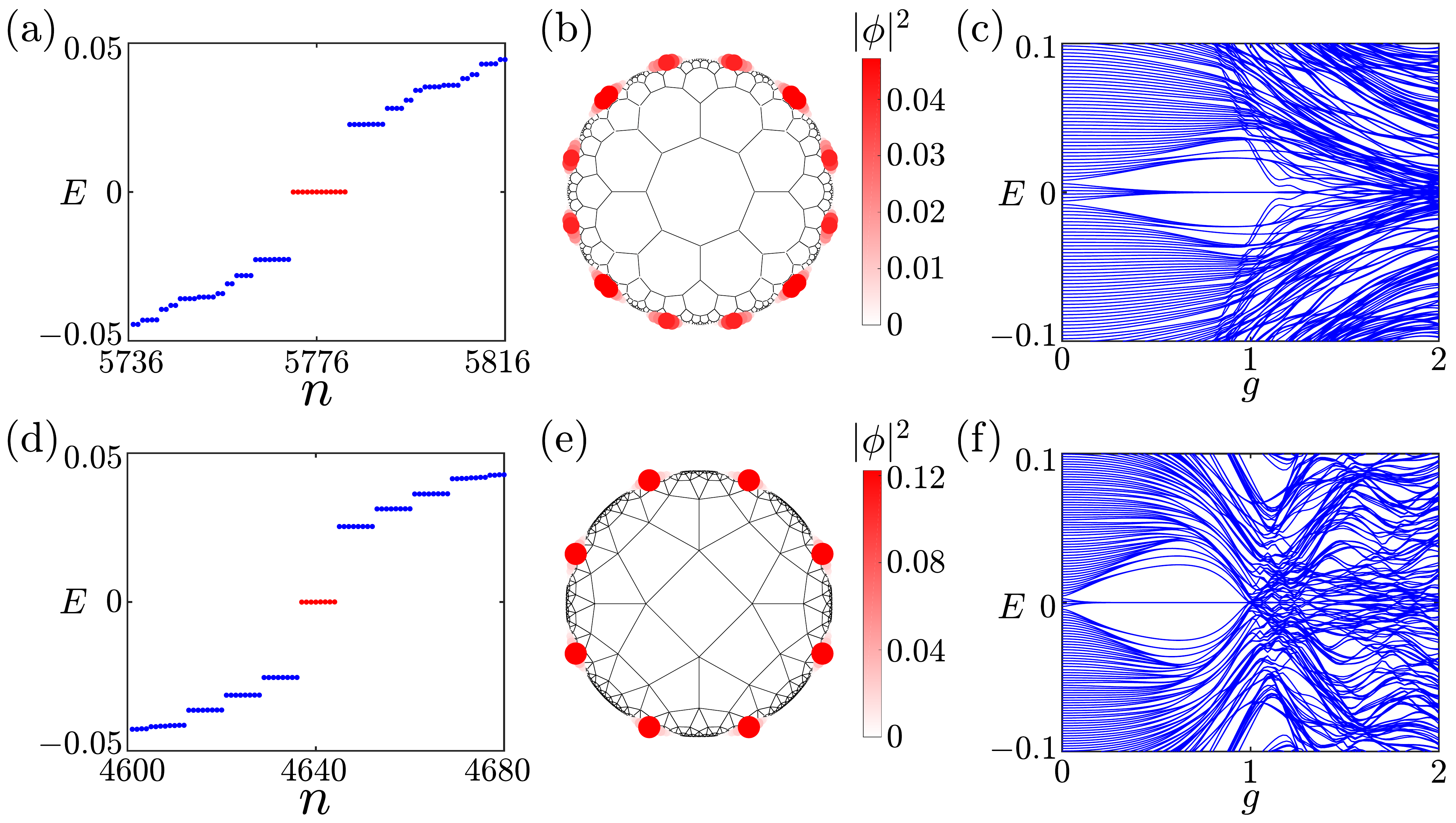} \caption{(a) Energy spectrum of Hamiltonian $H_{1}$ in the hyperbolic $\{8,3\}$ lattice. (b) The probability distribution of the 12 zero-energy modes marked with red dots in (a). We take the parameter $g=1$ in (a) and (b). (c) The energy spectrum of Hamiltonian $H_{1}$ in the hyperbolic $\{8,3\}$ lattice with respect to $g$. The parameters are $M=-1.2$, $t_{1}=t_{2}=1$, $\eta=6$, $N=2888$ in (a)--(c). (d) Energy spectrum of Hamiltonian $H_{1}$ in the hyperbolic $\{4,5\}$ lattice. (e) The probability distribution of the 8 zero-energy modes marked with red dots in (d). We take the parameter $g=0.6$ in (d) and (e). (f) The energy spectrum of Hamiltonian $H_{1}$ in the hyperbolic $\{4,5\}$ lattice with respect to $g$. The parameters are $M=-1.5$, $t_{1}=t_{2}=1$, $\eta=4$, $N=2320$ in (d)--(f).}%
	\label{figB1}
\end{figure}

\begin{figure}[t]
	\includegraphics[width=8.5cm]{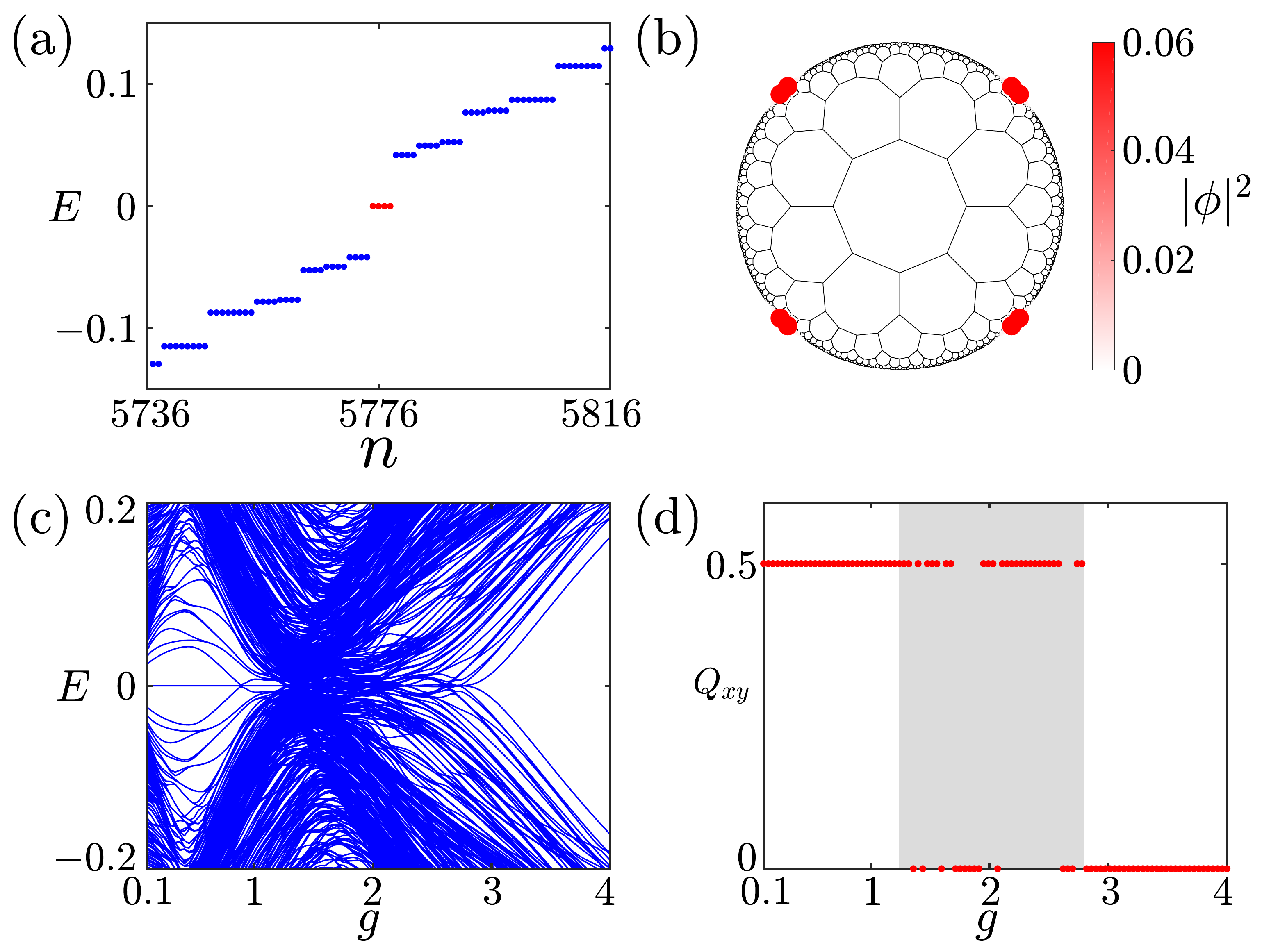} \caption{(a) Energy spectrum of Hamiltonian $H_{1}$ in the hyperbolic $\{8,3\}$ lattice. (b) The probability distribution of the four zero-energy modes marked with red dots in (a). We take the parameter $g=0.5$ in (a) and (b). (c) The energy spectrum of Hamiltonian $H_{1}$ in the hyperbolic $\{8,3\}$ lattice with respect to $g$. (d) The quadrupole moment $Q_{xy}$ as a function of $g$. The gray shaded area represents the area where the bulk energy gap is zero. Here, we take the parameters $M=-1.2$, $t_{1}=t_{2}=1$, $\eta=2$, and $N=2888$.}%
	\label{figB2}
\end{figure}

In Fig.~\ref{figB1}(a), we show the energy spectrum of Hamiltonian $H_{1}$ in the hyperbolic $\{8,3\}$ lattice when $\eta=6$. It is found that there are 12 zero energies marked with red dots in the energy gap. These zero energies originate from the mass domain walls constructed by the Wilson mass term at the edge. As shown in Fig.~\ref{figB1}(b), these zero-energy eigenstates are localized in zero dimensions at the edge of the finite hyperbolic lattice. To study the changes that occur in the system when the Wilson mass $g$ is changed, we plot the energy spectrum of Hamiltonian $H_{1}$ with respect to $g$ in Fig.~\ref{figB1}(c). We can find that a Wilson mass of suitable strength forces the edge of the finite hyperbolic lattice to produce the 0D mass domain walls, resulting in zero-energy effective corner states. These zero-energy effective corner states are destroyed when the strength of the Wilson mass exceeds a critical value. Without restricting to this case, we also observe a similar phenomenon by applying Hamiltonian $H_{1}$ with $\eta=4\neq p/2$ to the hyperbolic $\{4,5\}$ lattice as shown in Figs.~\ref{figB1}(d)--\ref{figB1}(f), and applying Hamiltonian $H_{1}$ with $\eta=2\neq p/2$ to the hyperbolic $\{8,3\}$ lattice as shown in Figs.~\ref{figB2}(a)--\ref{figB2}(c). In the main text, we present that the quadrupole moment can be used to characterize the HOTI phase of a hyperbolic lattice with four zero-energy states. Therefore, we plot the quadrupole moment of Hamiltonian $H_{1}$ in hyperbolic $\{8,3\}$ lattice as a function of Wilson mass $g$ when $\eta=2$ in Fig.~\ref{figB2}(d). In Fig.~\ref{figB2}(d), except for the shaded area, the platform with quadrupole moment $Q_{xy}=0.5$ represents the HOTI phase with four zero-energy states, while the platform with $Q_{xy}=0$ means that the system is in topological trivial phase. The quadrupole moment is not well-defined in a gapless system, so the quadrupole moment value in the shaded area cannot correctly represent the phase transition of the system.

Based on the symmetry analysis, we find that the Hamiltonian $H_{1}$ preserves only the PHS $P$ and the effective chiral symmetry $S m_{z}$ when $\eta\neq p/2$. And the two combined symmetries $C_{p}T$ and $C_{p}m_{z}$ associated with rotation symmetry are not preserved. These results suggest that the variation period of the Wilson mass determines the number of zero-energy effective corner states, and the Wilson mass term at $\eta\neq p/2$ is still able to induce the quantum spin Hall insulator to transform into a HOTI with $2\eta$ zero-energy effective corner states.

\section{Finite-size effect}
\label{AppendixFSE}

\begin{figure}[t]
	\includegraphics[width=8.5cm]{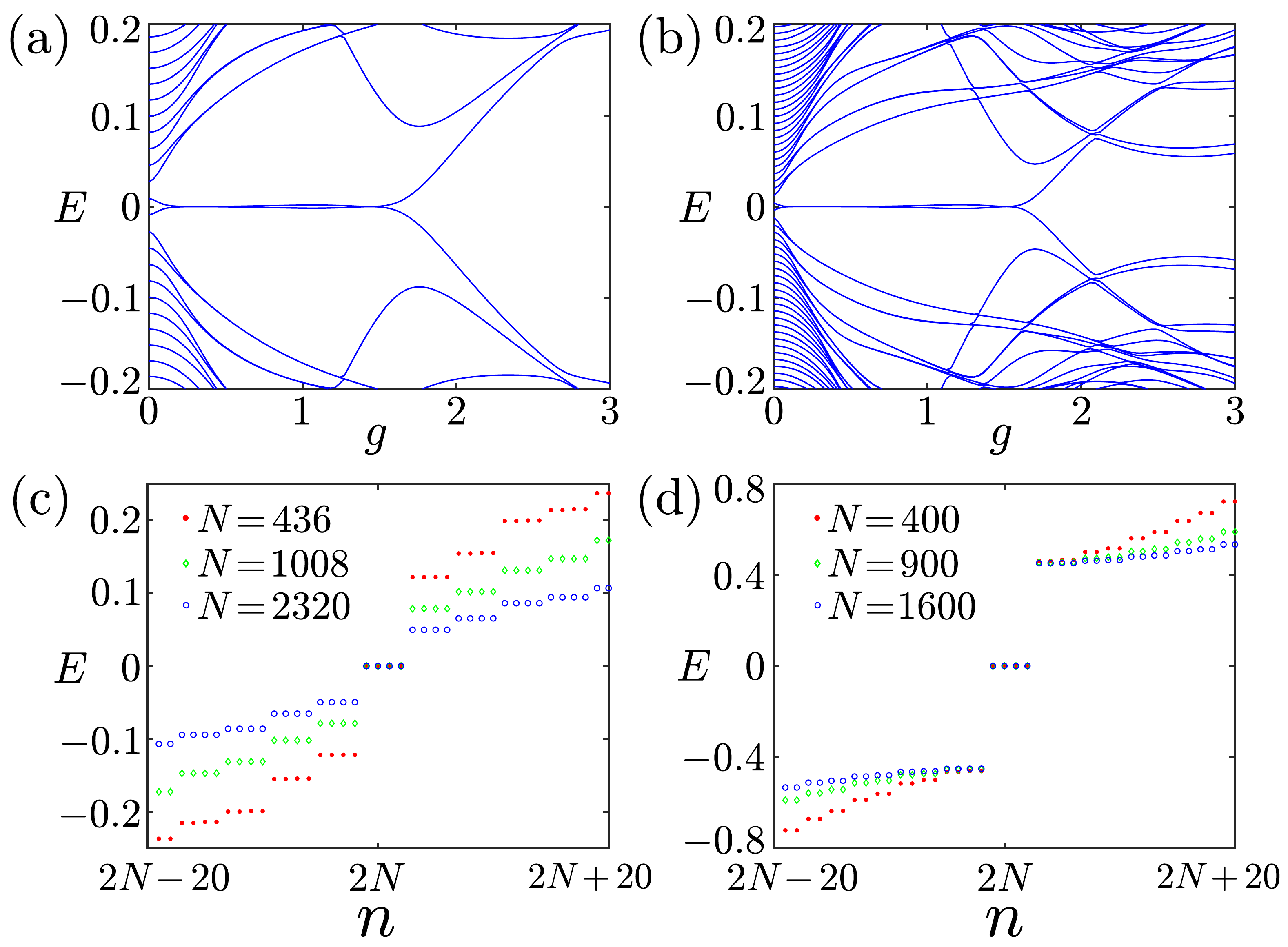} \caption{(a), (b) Energy spectrum of Hamiltonian $H_{1}$ in finite hyperbolic $\{4,5\}$ lattices of different sizes with respect to $g$. We take the number of sites $N=436$ for (a) and $N=1008$ for (b). (c) Energy spectrum of Hamiltonian $H_{1}$ in finite hyperbolic $\{4,5\}$ lattices of different sizes. The red dots, green diamonds, and blue circles represent hyperbolic $\{4,5\}$ lattices with the number of sites $N=436$, $N=1008$, and $N=2320$, respectively. (d) Energy spectrum of Hamiltonian $H_{1}$ in finite square lattices of different sizes. The red dots, green diamonds, and blue circles represent square samples with the number of sites $N=400$, $N=900$, and $N=1600$, respectively. Here, we take parameters $M=-1$, $t_{1}=t_{2}=1$, $g=0.5$, and $\eta=2$.}%
	\label{figC1}
\end{figure}

In this appendix, we explore the response of systems with zero-energy effective corner states to changes in sample size. In Figs.~\ref{figC1}(a) and \ref{figC1}(b), we show the energy spectrum of Hamiltonian $H_{1}$ in samples of different sizes with respect to Wilson mass $g$. In Fig.~\ref{figC1}(c), we show the energy spectrum of Hamiltonian $H_{1}$ in the hyperbolic $\{4,5\}$ lattice, with the colors of the dots representing samples of different sizes. The energy gap where the zero-energy effective corner states are located decreases as the size of the sample increases. To compare the results in the Euclidean system, we take the example of the square lattice. In Fig.~\ref{figC1}(d), we show the energy spectrum of Hamiltonian $H_{1}$ in the square lattice, with the colors of the dots representing square samples of different sizes. We find that when the sample size changes, the energy gap of square lattice in Euclidean system is stable, while the energy gap of hyperbolic $\{4,5\}$ lattice decreases with the increase of sample size. This is because the edge of the finite hyperbolic lattice is not smooth and there are bonds with different orientations on the edge. These bonds lead to the appearance of in-gap local states, and the number of such bonds increases with the increase in the size of the finite hyperbolic lattice, thus the edge energy gap decreases gradually. A similar phenomenon has been reported in the HOTIs on a fractal lattice, where the energy gap also decreases as the system size increases~\cite{PhysRevB.105.L201301}. However, this gap remains finite as the sample size approaches the thermodynamic limit. Due to limitations in computational power, it is currently unclear whether the edge gap in hyperbolic lattices remains finite in the thermodynamic limit, and this will be investigated in our future works.

\section{Higher-order hyperbolic topological insulators with time-reversal symmetry}
\label{AppendixQI}

In general, TRS breaking is not necessary for HOTIs. Many works on TRS-preserving HOTIs have been proposed~\cite{doi:10.1126/sciadv.aat0346, PhysRevLett.119.246401, Schindler2018, PhysRevX.8.031070, doi:10.1126/science.aah6442, PhysRevLett.119.246402, PhysRevLett.123.256402, PhysRevB.96.245115, Lee2020, PhysRevB.97.205135, PhysRevB.97.205136, PhysRevX.9.011012, PhysRevLett.124.036803, PhysRevB.99.245151, PhysRevB.102.241102, PhysRevB.104.245302, PhysRevB.98.201114, Noh2018}. Here we apply TRS-preserving quadrupole insulator (QI) model to the hyperbolic lattice~\cite{doi:10.1126/science.aah6442}. The QI model in the real space can be described by the following tight-binding model Hamiltonian~\cite{PhysRevLett.124.036803}:
\begin{align}
H_{\rm{QI}}&=\gamma\sum_{j}c_{j}^{\dagger}(\Gamma_{2}+\Gamma_{4})c_{j}\nonumber\\
&+\frac{\lambda}{2}\sum_{\left <j,k \right >}c_{j}^{\dagger}[ |\cos(\theta_{jk})|\Gamma_{4}-i \cos(\theta_{jk})\Gamma_{3}\nonumber\\
&+|\sin(\theta_{jk})|\Gamma_{2}-i \sin(\theta_{jk})\Gamma_{1}] c_{k},
\label{QI}
\end{align}
where $c_{j}^{\dagger}$ and $c_{j}$ are the creation and annihilation operators of electrons on site $j$. $\theta_{jk}$ represents the polar angle of the vector from the site $k$ to the site $j$ in the Poincar\'{e} disk. The vector from the site $k$ to the site $j$ mentioned here refers to the vector composed of two points in the Euclidean plane, not the geodesic line in the Poincar\'{e} disk model. $\Gamma_{1}=-\sigma_{y}\sigma_{x}$, $\Gamma_{2}=-\sigma_{y}\sigma_{y}$, $\Gamma_{3}=-\sigma_{y}\sigma_{z}$, and $\Gamma_{4}=\sigma_{x}\sigma_{0}$. $\sigma_{x,y,z}$ are the Pauli matrices acting on the sublattice, $\sigma_{0}$ is the identity matrix. $\gamma$ represents the hopping amplitude between the sublattices of the same site. $\lambda$ represents the hopping amplitude between the nearest-neighbor sites. The nearest-neighbor distance here refers to the nearest hyperbolic distance between two sites in the Poincar\'{e} disk ~\cite{PhysRevB.105.125118}. The Hamiltonian $H_{\rm QI}$ respects TRS $T=\mathcal{K}$, PHS $P=\sigma_{z}\sigma_{0}\mathcal{K}$, and chiral symmetry $S=PT=\sigma_{z}\sigma_{0}$.

\begin{figure}[t]
	\includegraphics[width=8.5cm]{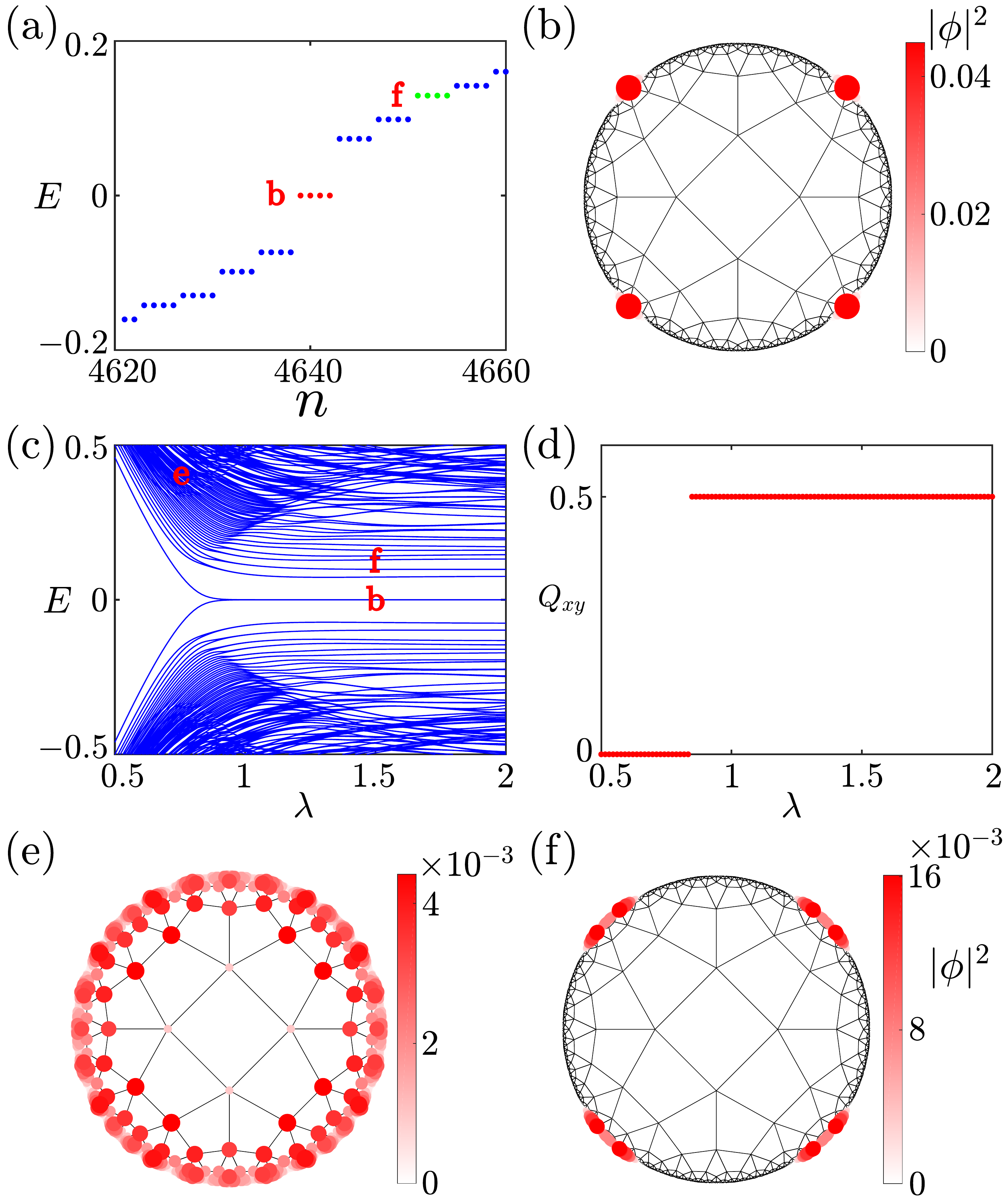} \caption{(a) Energy spectrum of Hamiltonian $H_{\rm{QI}}$ in the hyperbolic $\{4,5\}$ lattice. (b) The probability distribution of the four zero-energy modes [the red dots in (a)]. We take the parameter $\lambda=1.5$ in (a) and (b). (c) The energy spectrum of Hamiltonian $H_{\rm{QI}}$ in the hyperbolic $\{4,5\}$ lattice with respect to $\lambda$. (d) The quadrupole moment $Q_{xy}$ as a function of $\lambda$. (e) The probability distribution of the bulk states ($E=0.3972$) when $\lambda=0.75$. (f) The probability distribution of the local states ($E=0.1293$) [the green dots in (a)]. Here, we take the parameters $\gamma=1$, and $N=2320$.}%
	\label{figD1}
\end{figure}

In Fig.~\ref{figD1}(a), we show the energy spectrum of Hamiltonian $H_{\rm QI}$ in the hyperbolic $\{4,5\}$ lattice. It is found that there are four degenerate zero energies, which are similar to the results in the square lattice and quasicrystals~\cite{doi:10.1126/science.aah6442, PhysRevB.96.245115, PhysRevLett.124.036803, PhysRevB.104.245302}. These zero-energy eigenstates are localized in 0D form at the edge of the hyperbolic lattice as shown in Fig.~\ref{figD1}(b). In the square lattice, the zero-energy corner states are localized at the four vertices of the square sample when the inter-cell hopping amplitude is greater than the intra-cell hopping amplitude~\cite{doi:10.1126/science.aah6442}. Similar to the case of the square lattice, this system also requires certain conditions to be satisfied in the hyperbolic lattice to exhibit the zero-energy effective corner states. Therefore, we show in Fig.~\ref{figD1}(c) the energy spectrum of Hamiltonian $H_{\rm{QI}}$ in the hyperbolic $\{4,5\}$ lattice as a function of $\lambda$. It is obvious that four degenerate zero-energy states appear in the system only when $\lambda>0.9242$. This model with polarization angle $\theta_{jk}$ does not possess translation symmetry in the hyperbolic lattice, thus we use real-space quadrupole moment as a topological invariant to characterize the hyperbolic higher-order topology. In Fig.~\ref{figD1}(d), we show the quadrupole moment $Q_{xy}$ as a function of $\lambda$. Then, the system transitions from a normal insulator phase to a HOTI phase with four zero-energy effective corner states when $\lambda$ exceeds the critical value.

Furthermore, in the spectrum shown in Fig.~\ref{figD1}(c), one clearly recognizes regions in the $(\lambda,E)$ plane with ``more dense'' and ``less dense'' distribution of eigenenergies. The ``more dense'' region corresponds to the bulk band of states~[Fig.~\ref{figD1}(e)], whereas the ``less dense'' region corresponds to the edge band of states~[Fig.~\ref{figD1}(f)], which arises from bonds of a variety of orientations at the edge of the finite hyperbolic lattice. Then, the phase at $\lambda>0.9$ (where the corner modes are observed), would likely be interpreted as having metallic edge states. This seems to provide further evidence for the conclusion that the finite edge gap may be a result of the finite-size effect and that the edge gap may drop to zero in the thermodynamic limit, which is consistent with the discussions in Appendix~\ref{AppendixFSE}.

In this appendix, we apply the QI model in real space to the hyperbolic $\{4,5\}$ lattice that preserves TRS. In this system, we observe four zero-energy effective corner states. Numerical calculations of the real-space quadrupole moment show that these zero-energy effective corner states are topologically non-trivial.

\newpage
\bibliographystyle{apsrev4-1-etal-title_6authors}
%\bibliography{bibfile}
%
\end{document}